\pdfoutput=1

\documentclass[11pt]{article}

\usepackage[final]{acl}
\usepackage{tcolorbox}

\usepackage{times}
\usepackage{latexsym}
\usepackage{amsmath}
\usepackage{algorithm}
\usepackage{algorithmic}
\usepackage{amsfonts}  
\usepackage{subcaption}
\usepackage{multirow}
\usepackage[T1]{fontenc}

\usepackage[utf8]{inputenc}

\usepackage{microtype}

\usepackage{inconsolata}

\usepackage{graphicx}

\usepackage{booktabs}

\title{RecBase: Generative Foundation Model Pretraining  for Zero-Shot Recommendation}

\author{
  \textbf{Sashuai Zhou\textsuperscript{1,3}},
  \textbf{Weinan Gan\textsuperscript{2}},
  \textbf{Qijiong Liu\textsuperscript{4}},
  \textbf{Ke Lei\textsuperscript{1}},
  \textbf{Jieming Zhu\textsuperscript{2}\footnotemark[2]},\\
  \textbf{Hai Huang\textsuperscript{1}},
  \textbf{Yan Xia\textsuperscript{1}},
  \textbf{Ruiming Tang\textsuperscript{2}},
  \textbf{Zhenhua Dong\textsuperscript{2}},
   \textbf{Zhou Zhao\textsuperscript{1,3}\footnotemark[2]}
\\
 \textsuperscript{1}Zhejiang University \quad
 \textsuperscript{2}Huawei Noah’s Ark Lab \quad 
 \textsuperscript{3}Shanghai AI Lab \quad 
 \textsuperscript{4}The HK PolyU
\\
\href{mailto:zhousashuai@zju.edu.cn}{zhousashuai@zju.edu.cn} \quad
  \href{mailto:jiemingzhu@ieee.org}{jiemingzhu@ieee.org} \quad
  \href{mailto:zhaozhou@zju.edu.cn}{zhaozhou@zju.edu.cn}
}

\begin{document}
\maketitle

\footnotetext[2]{Corresponding Authors.}

\begin{abstract}

Recent advances in LLM-based recommendation have shown promise, yet their cross-domain generalization is hindered by a fundamental mismatch between language-centric pretraining and the recommendation task. Existing methods, relying on language-level knowledge, fail to capture dynamic, item-level user interests across domains. To bridge this gap, we propose RecBase, a domain-agnostic foundational model pretrained with a recommendation-oriented objective. RecBase leverages a large-scale, heterogeneous, cross-domain corpus with unified textual representations and feature mappings to enhance cross-domain generalization. To further align item semantics across domains, we introduce a unified item tokenizer that encodes items into hierarchical concept identifiers, enabling structured representation and efficient vocabulary sharing. The model is trained using an autoregressive objective to capture complex item-level sequential patterns. On eight real-world datasets, our 1.5B-parameter model matches or surpasses the performance of LLM baselines up to 7B parameters in zero-shot and cross-domain recommendation tasks.

\end{abstract}

\section{Introduction}
In recent years, large language models (LLMs)~\cite{LLMSurvey} have demonstrated powerful capabilities in zero-shot learning, multi-task unification, and multi-domain generalization. Inspired by these successes, an intriguing yet underexplored area is the development of foundational models specifically tailored for recommender systems~\cite{FoundationModelSurvey,MMRecSurvey}. Recommender systems \cite{BARS} are essential for helping users discover content of interest and have been widely applied across various domains, such as videos, music, and products. An ideal foundation model for recommender systems should be capable of addressing diverse recommendation tasks across different domains while also performing  effectively in zero-shot and few-shot (e.g., cold-start) settings.

Toward this goal, existing research aims to leverage the strengths of LLMs to enhance the effectiveness and versatility of recommender systems, a field referred to as LLM-based recommendation \cite{LLM4RecSurvey}. For instance, some initial efforts, such as those described in \cite{ChatGPT4Rec, ZeroShotRec, ZeroShotRank}, explore the direct application of LLMs for zero-shot recommendation. Other works, like P5 \cite{P5} and GenRec \cite{Genrec2023}, focus on continuing pretraining of LLMs in multi-domain and multi-task recommendation settings. Additionally, some other studies such as \cite{petrov_et_al_2023GPTrec, tallrec, ClickPrompt} investigate efficient fine-tuning and alignment of LLMs for downstream recommendation tasks. However, these approaches face several limitations: 1) Input Representation: Recommendation data often needs to be mapped into language modalities, which may not effectively represent user sequences as shown in our experiments. 2) Knowledge Gap: The knowledge gap between language models and recommendation tasks makes them struggle with modeling item-item co-relationships, hindering their performance in zero-shot recommendations. 3) Model Alignment: Fine-tuning language models to align with recommendation models using downstream task datasets can compromise the model's ability to effectively handle zero-shot and cross-domain recommendations. As a result, such approaches often fail to meet the expectations set for foundational recommendation models.

In this paper, we aim to bridge the gap by making the first effort to pretrain a foundational model from scratch (dubbed RecBase), supporting both zero-shot and multi-domain recommendation settings. To achieve this, we leverage LLMs solely as encoders for unified semantic representation and then model item-item relationships through generative pretraining on large-scale, open-domain, recommendation-oriented item sequence data. Specifically, our work makes the following technical contributions: \textbf{1) Data Collection and Representation}: We compile a large-scale, open-domain recommendation dataset spanning 15 different domains (comprising 4.5M items and 35M interactions). We uniformly extract textual representations of items to serve as a data source for pretraining across various domains. \textbf{2) Unified Item Tokenizer}: Instead of relying on ID-based sequence modeling, which lacks semantics, or language-based modeling, which is often verbose, we propose a general item tokenizer that unifies item representations across domains. Each item is tokenized into multi-level concept IDs, learned in a coarse-to-fine manner inspired by curriculum learning. This hierarchical encoding facilitates semantic alignment, reduces vocabulary size, and enables effective knowledge transfer across diverse domains. \textbf{3) Autoregressive Pretraining}: We adopt an autoregressive modeling paradigm for pretraining, where the model predicts the next token in a sequence. This approach enables learning item co-relationships within a unified concept token space, thereby enhancing the model’s generalization in zero-shot and cross-domain settings. For clarity and consistency, we provide two pretrained versions of our foundation models: RecBase-0.3B and RecBase-1.5B.

To assess the model’s generalization ability in zero-shot and multi-domain settings, we design an evaluation framework that predicts, i.e., ranks, items users are likely to engage with across a wide range of recommendation tasks. We conduct extensive experiments on eight diverse and previously unseen datasets to examine the zero-shot and cross-domain performance of RecBase. The results show that our recommendation-oriented pretraining strategy substantially enhances the model’s ability to generalize to new domains without task-specific tuning. While fine-tuning on in-domain data can further improve performance, our primary emphasis in this work is on zero-shot generalization. In these settings, RecBase consistently outperforms language-model-based baselines, underscoring the effectiveness of recommendation-aligned pretraining and its potential to support robust and adaptable recommendation across domains. RecBase will be open-sourced at \href{https://github.com/reczoo/RecBase}{https://github.com/reczoo/RecBase}.

\begin{figure*}[h]
    \centering
    \includegraphics[width=\textwidth]{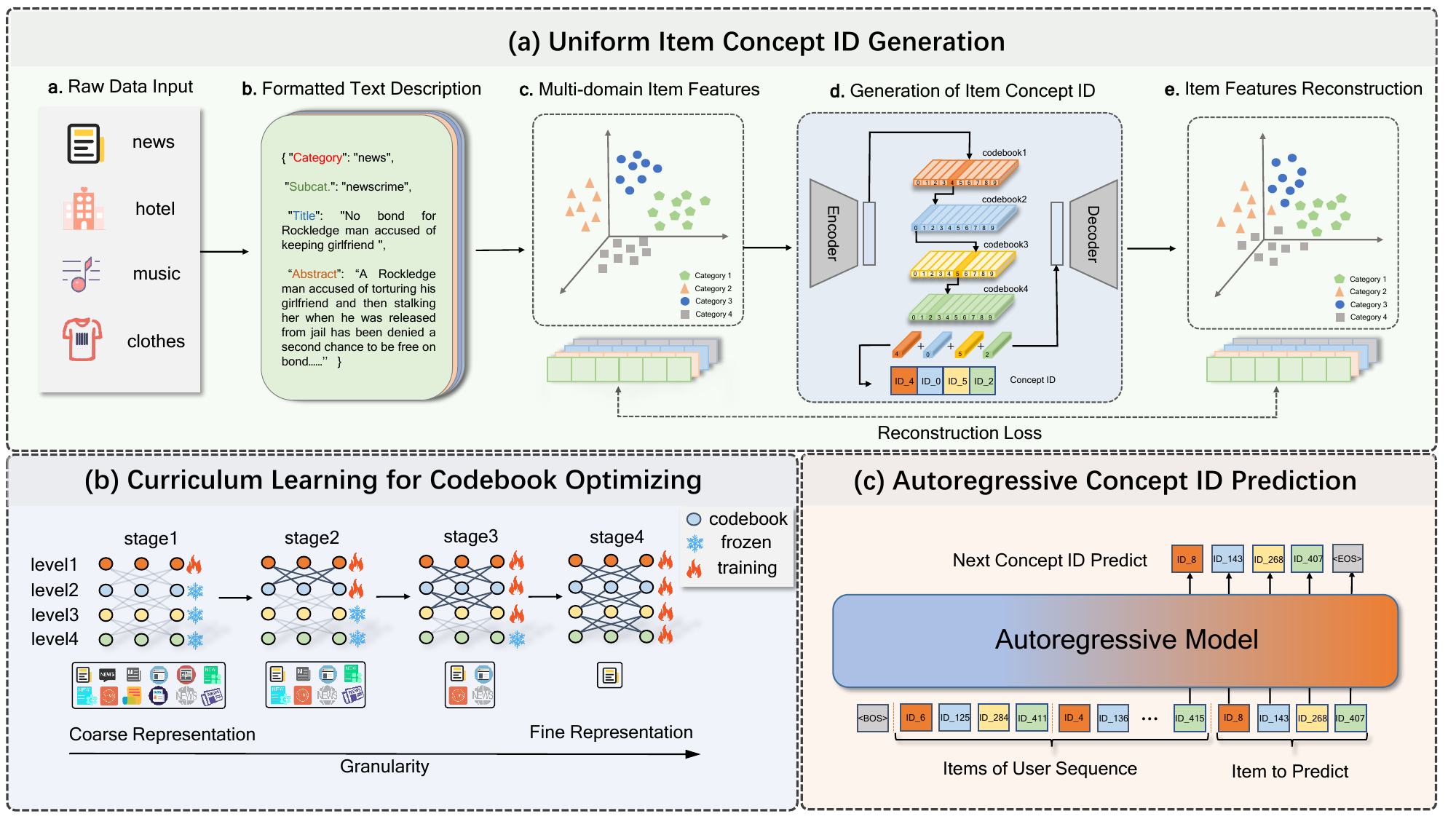}  
    \caption{Overview of the RecBase model. (a) illustrates the use of discrete representation techniques to transform product descriptions from multiple domains into unified concept ID sequences. (b) depicts the curriculum learning process for optimizing codebook learning. (c) demonstrates how the autoregressive model leverages discretized concept IDs to predict the next item in the sequence, effectively capturing item relationships for recommendation.}  
    \label{fig:example}  
\end{figure*}

\section{Related Work}

\subsection{LLM-based Recommendation}

Recent advancements in LLM-based recommendation systems have focused on both item scoring and generation tasks. In item scoring, models like M6-Rec \cite{M6-rec}, Prompt4NR \cite{prompt4nr}, TabLLM \cite{tabllm}, and TALLRec \cite{tallrec} transform user-item data into natural language representations, either generating item descriptions for scoring or reframing the task as a cloze-style prediction. Similarly, ONCE \citep{liu_et_al_2023_afirst} offers a generative framework for content-based recommendation, and CLLM4Rec \citep{zhu_et_al_2024collac} enhances collaboration in LLM-based systems. For item generation, approaches like GPT4Rec \citep{petrov_et_al_2023GPTrec}, P5 \cite{P5}, EAGER~\cite{EAGER}, and EAGER-LLM~\cite{EAGER-LLM} leverage generative models to predict the next item based on user behavior, with DiffuRec incorporating uncertainty into sequential recommendations. Additionally, GIRL \citep{zheng_et_al_2023GBR} demonstrates how LLMs can improve job recommendations. While LLM-based methods have advanced recommendations, they struggle with the semantic gap between language-based representations and structured recommendation data. Based on this, we propose a pretrained model that directly learns from recommendation-specific representations to bridge the semantic gap and enhance adaptability across domains.

\subsection{Item Representation for Recommendation}

Recent works have made significant strides in enhancing recommendation systems by improving item representations. Hierarchical models, such as those using graph neural networks to aggregate item information into representations \cite{HierarchicalFashion,HierarchicalAttentive}, have been shown to refine user profiles and capture dependencies within and across interactions. Techniques like cross-view contrastive learning \cite{crosscbr} have also been proposed to model user-bundle and user-item interactions, facilitating better generalization across domains. Additionally, methods addressing sequential recommendation challenges have focused on mitigating item representation divergence \cite{RAM}, further enhancing learning efficiency. Another noteworthy advancement is the use of Semantic IDs for items \cite{tiger,LCrec,colarec,CoST}, where generative retrieval frameworks predict the next item in a sequence, enhancing generalization, particularly in zero-shot scenarios. While these methods show progress, they are often constrained by domain-specific data, limiting their ability to generalize across different contexts. Our approach aims to overcome these limitations by constructing a unified feature representation that enables better generalization across diverse domains.

\section{Method}


Our proposed method consists of two main stages. First, item representations are mapped into a unified discretized space, generating a concept ID sequence for each item (Section 3.2). Second, this concept ID sequence is used to autoregressively predict the next in the sequence (Section 3.3).

\subsection{Preliminary}
To facilitate the learning of user behavior based on item history by large language models, it is necessary to discretize continuous item semantic embeddings into unified discrete tokens. This can be achieved through Residual Quantized Variational Autoencoder (RQ-VAE) \cite{RQ-VAe}. Given an input $e \in \mathbb{R}^d$, RQ-VAE first encodes $e$ into the latent space using an encoder $\mathcal{E}$:

\begin{equation}
    z := \mathcal{E}(e)
\end{equation}

RQ-VAE extends Vector Quantized Variational Autoencoder (VQ-VAE) \cite{Vq-vae} by introducing the concept of hierarchical quantization, which progressively quantizes the latent representation $z$. Specifically, at each level $d$, the residual $r_d$ is quantized by mapping it to the nearest embedding $e_{c_d}$ in the level-specific codebook $C_d := {e_k}_{k=1}^K$, where

\begin{equation}
    c_d = \arg \min_k \|r_d - e_k\|
\end{equation}

The residual for the next level is computed as

\begin{equation}
    r_{d+1} := r_d - e_{c_d}
\end{equation}

This process is recursively repeated $m$ times to generate a tuple of $m$ codewords representing the Semantic ID, approximating the input from coarse to fine granularity. Separate codebooks for each level allow for varying granularities as the residual norms decrease.

\begin{algorithm}[ht]
\renewcommand{\algorithmicrequire}{\textbf{Input:}}
\renewcommand{\algorithmicensure}{\textbf{Output:}}
\caption{Training Process of CL-VAE}
\label{alg:rqvae_training}
\begin{algorithmic}[1]
\REQUIRE semantic embeddings $e \in \mathbb{R}^{B \times D}$,
encoder $\mathcal{E}$, decoder $\mathcal{D}$,
m-level codebooks $\{C_d\}_{d=1}^m$


\STATE \text{used\_level}=0
\STATE \textbf{Encoding:} 
\STATE $\mu, \text{log}\sigma^2 \gets \mathcal{E}(e)$ 
\STATE $z \gets \mu + \sigma \odot \epsilon$ \COMMENT{Sampling $\epsilon \sim \mathcal{N}(0, I)$}

\STATE \textbf{Curriculum Learning:} 
\IF{Total loss converges}
    \STATE $\text{used\_level} \gets \text{used\_level} + 1$ 
\ENDIF

\STATE \textbf{Quantization:} 
\STATE $z_q, \text{indices}, L_Q \gets \text{rqvae}(z, \{C_d\}_{d=0}^{\text{used\_level}})$

\STATE \textbf{Codebook Initialization:} 
\IF{codebook usage rate is low}
    \STATE $C_0 \gets \text{KMeans}(z)$
\ENDIF

\STATE \textbf{Loss Computation:} 
\STATE $L_e \gets \text{entropy\_loss}(\text{indices})$ 
\STATE $e_{\text{recon}} \gets \mathcal{D}(z_q)$ 
\STATE $L_{\text{R}} \gets \text{MSE}(e_{\text{recon}}, e)$ 
\STATE $L_{total} \gets L_{\text{R}} + L_Q + \gamma L_e$ 

\end{algorithmic}
\end{algorithm}

\subsection{Unified Feature Representation Space }
To ensure a consistent and structured representation of items across different domains, we first standardize item descriptions into a unified format. This allows for uniform processing and facilitates transformation into feature embeddings using a shared encoder. By maintaining structural consistency, the model can effectively learn and compare items across domains within a common feature space.

For a highly generalizable large language recommendation model, it is essential that IDs derived from item embeddings are evenly distributed across the ID space. This ensures that the representations of items from diverse domains are well spread out, making it easier for the model to generalize across various categories. However, during RQ-VAE training, codebook collapse often occurs, where the majority of inputs, regardless of their domain, are mapped to only a small subset of codebook vectors. This results in a situation where a new item, especially one from a previously unseen domain, may be encoded into a token that the large language model (LLM) has never encountered, leading to suboptimal performance. To mitigate this issue, we propose \textbf{Curriculum Learning Enhanced RQ-VAE (CL-VAE)}, which improves the model's robustness and enhances its ability to handle zero-shot scenarios across diverse domains.

As shown in Figure~\ref{fig:example}, in CL-VAE, we introduce curriculum learning \cite{curriculum,curri2}, where the core idea is to allow the model to progressively learn tasks from simple to complex~\cite{curri3}, rather than directly learning complex tasks. We observed that the hierarchical structure of RQ-VAE inherently aligns with the principles of curriculum learning. We perform staged training for the different levels of the codebook: initially, we train the first layer for \( n \) epochs to allow it to fully learn basic feature representations, and after the loss stabilizes, we add the second layer for further training, and so on. This staged training approach not only reduces the complexity of initial training but also enhances the model's convergence stability. By gradually building hierarchical representations, CL-VAE effectively maps item representations from diverse domains into a unified concept ID space, enabling the large language model to generalize more effectively across different distributions.
\begin{figure}[t]
    \centering
    \begin{subfigure}{0.492\columnwidth} 
        \centering
        \includegraphics[width=\columnwidth]{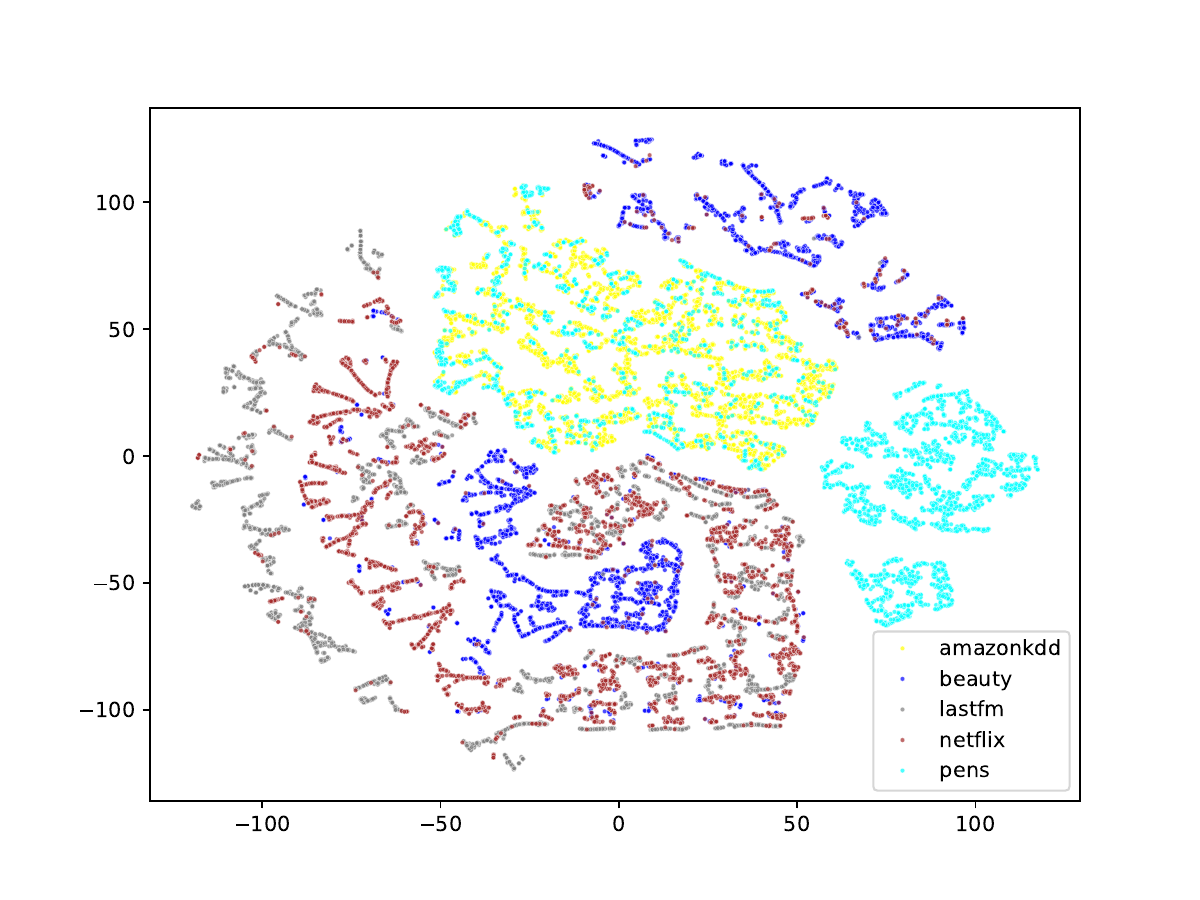}
        \caption{RQ-VAE}
        \label{fig:cluster1}
    \end{subfigure}
    \hfill
    \begin{subfigure}{0.492\columnwidth} %
        \centering
        \includegraphics[width=\columnwidth]{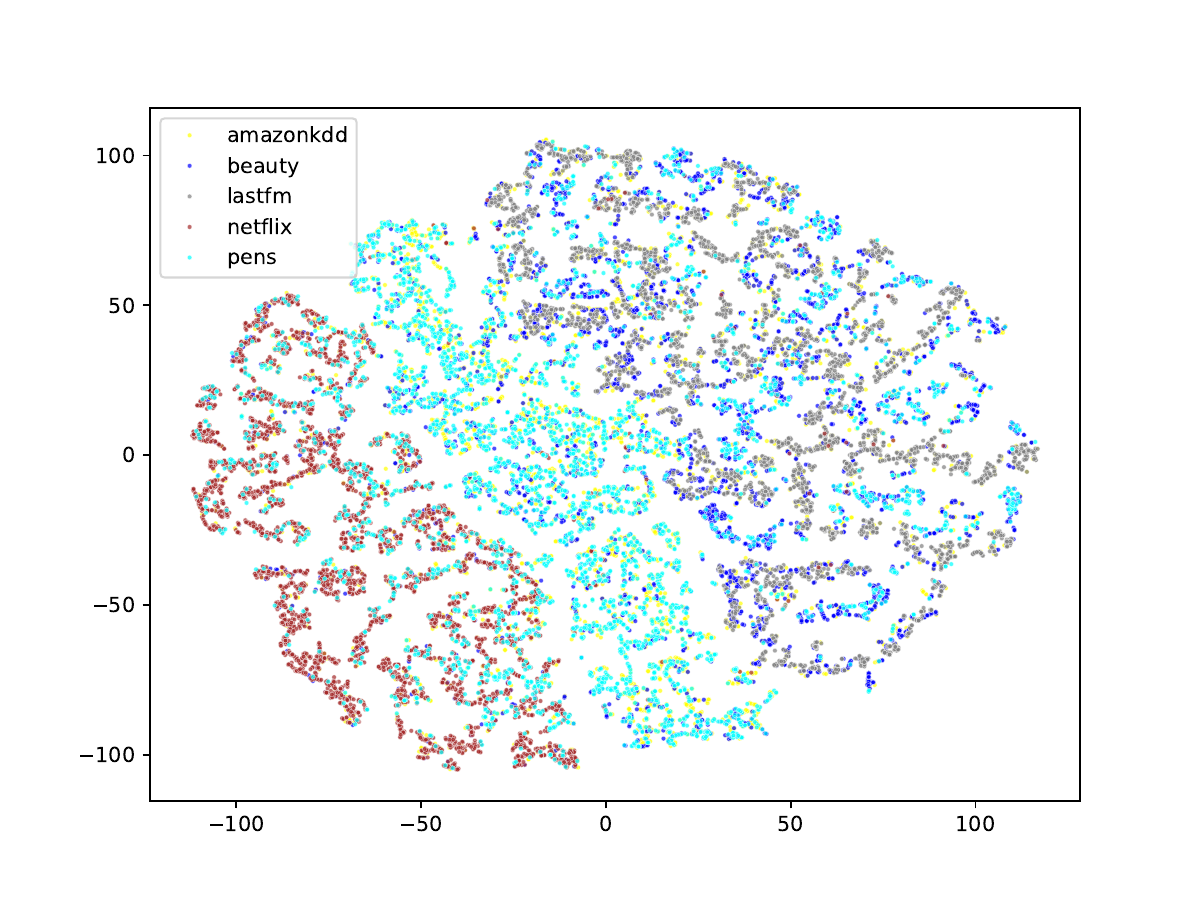}
        \caption{CL-VAE}
        \label{fig:cluster2}
    \end{subfigure}
    \caption{Visualization of t-SNE Clustering in the ID Space: (a) Discrete ID space learned by RQ-VAE, (b)Discrete ID space learned by CL-VAE.}

    \label{fig:clustering}
\end{figure}

To further mitigate codebook collapse, particularly in cases where certain codebook vectors are sparsely utilized, we determine whether to reinitialize the first-level codebook based on its usage during training. This reinitialization provides a new optimization starting point for the first-level codebook, ensuring sufficient learning of low-level features and preventing collapse, thereby enhancing the overall performance of the model. 

The loss for the modified model is composed of reconstruction loss $\mathcal{L}_{\text{R}}$, codebook loss and commitment loss $\mathcal{L}_{\text{Q}}$, entropy loss $\mathcal{L}_{\text{E}}$:

\begin{equation}
\mathcal{L}(x) := \mathcal{L}_{\text{R}} + \mathcal{L}_{\text{Q}} + \gamma\mathcal{L}_{\text{E}},
\end{equation}

\begin{equation}
\mathcal{L}_{\text{R}} := |x - \hat{x}|^2,
\end{equation}
\begin{equation}
\mathcal{L}_{\text{Q}} := \sum_{d=0}^{m-1} |\operatorname{sg}[r_i] - e_{c_i}|^2 + \beta |r_i - \operatorname{sg}[e_{c_i}]|^2,
\end{equation}
\begin{equation}
\mathcal{L}_{\text{E}} := -\sum^{m-1}_{d=0} \sum^K_{j=1} p_{d,j} \log p_{d,j}.
\end{equation}
here $\hat{x}$ is the output of the decoder, $\operatorname{sg}$ denotes the stop-gradient operation, and $p_{d,j}$ represents the usage frequency of codebook vector $e_j$ at level $d$. The first two loss terms are intrinsic to RQ-VAE, while the third term is an entropy penalty designed to promote more diverse utilization of the codebook. Together, these loss terms facilitate the joint training of the encoder, decoder, and codebook.

\begin{figure}[t]
    \centering
    \includegraphics[width=\columnwidth]{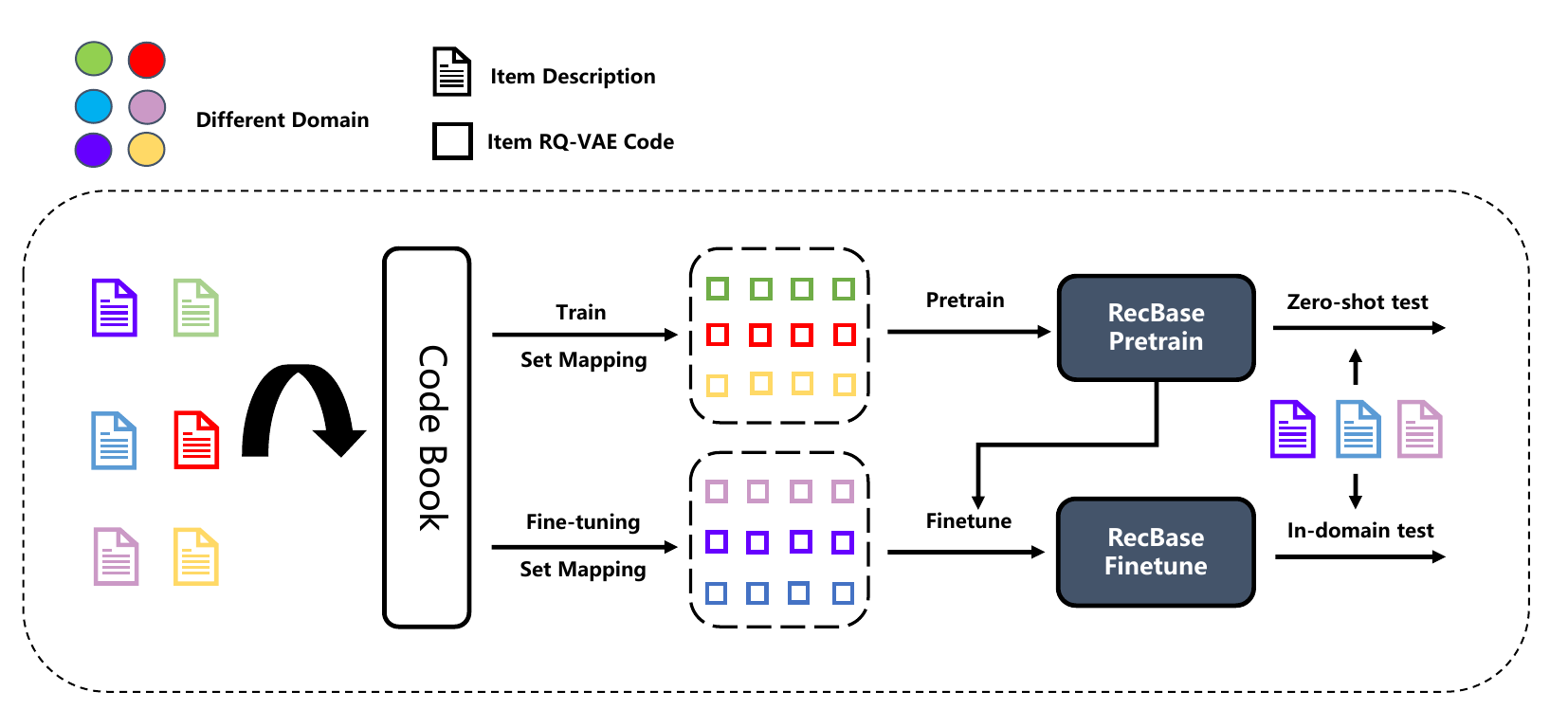}  %
\caption{Illustration of the zero-shot transfer and domain-specific fine-tuning process. The diagram shows how the pretrained model is tested in a zero-shot setting and fine-tuned for in-domain performance.}
    \label{fig:pipe}  %
\end{figure}

\subsection{Autoregressive Modeling}
After applying the unified discretization method, we obtain a sequence of concept IDs, where each item is represented as an \( m \)-bit semantic ID. To utilize this representation for recommendation, we convert a user's interaction history into a sequence of concept IDs, preserving the chronological order. Each item in the sequence is denoted as \( s_i = (s_i^1, s_i^2, \dots, s_i^m) \), where \( s_i^j \) is the \( j \)-th bit of the item's concept ID.  

Given a sequence of historical interactions \( S = (s_1, s_2, \dots, s_n) \), we train an autoregressive model to predict the next ID. The model takes the preceding sequence \( S_{<t} \) as input and outputs a probability distribution over each bit of the next ID \( s_t \):  

\begin{equation}  
P(s_t | S_{<t}) = \prod_{j=1}^m P(s_t^j | s_t^{<j}, S_{<t}),  
\end{equation}  
where \( P(s_t^j | s_t^{<j}, S_{<t}) \) is the probability of the \( j \)-th bit given its previous bits and the interaction history.  
Training is performed using the negative log-likelihood loss:  
\begin{equation}  
\mathcal{L} = -\sum_{t=1}^n \sum_{j=1}^m \log P(s_t^{j*} | s_t^{<j*}, S_{<t}),  
\end{equation}  
where \( s_t^{j*} \) is the ground truth bit value.  

During inference, the model generates the next item's concept ID bit by bit, treating concept IDs as tokens in its vocabulary. This structured representation enables the model to effectively capture user behavior patterns, leading to more accurate and diverse recommendations.

\begin{table*}[ht]
\centering
\renewcommand{\arraystretch}{1.2} 

\label{tab:main}
\caption{Zero-shot recommendation evaluation across multi-domain datasets.}

\resizebox{\textwidth}{!}{
\begin{tabular}{lcccccccccc}
\toprule
  &Size(M)& MIND & MovieLens & MicroLens & Goodreads & Yelp & Steam & H\&M & HotelRec & Overall \\
\midrule
P5  &223& 
0.4911 & 0.5138 & 0.5017 & 0.5027 & 0.5080 & 0.5296 & 0.4845 & 0.4905 & 0.5027\\
RecGPT  & 6,649& 
0.5078 & 0.5069 & 0.4703 & 0.5083 & 0.5140 & 0.4924 & 0.4875 & 0.4937 & 0.4976\\

BERT$_\text{base}$  &110&
0.4963 & 0.4934 & 0.4992 & 0.4958 & 0.4914 & 0.5002 & 0.5204 & 0.4955 & 0.4990\\
OPT$_\text{base}$  &331&
0.5490 & 0.5104 & 0.4773 & 0.5015 & 0.5158 & 0.4257 & 0.4555 & 0.5028 & 0.4922\\
OPT$_\text{large}$  &1,316&
0.5338 & 0.5174 & 0.5236 & 0.5042 & 0.5026 & 0.3825 & 0.5650 & 0.5026 & 0.5039\\
Qwen-2  &494& 
0.4886& 0.5138& 0.5701& 0.5148& 0.5077& 0.6399& 0.6287& 0.5311& 0.5493\\
Phi-2  &2,780& 
0.4851 & 0.5296 & 0.5078 & 0.5049 & 0.5186 & 0.6061 & 0.5447 & 0.4986 & 0.5244\\
Llama-2  &6,738& 
0.4945 & 0.6030 & 0.4877 & 0.5273 & \textbf{0.5378} & 0.5622 & 0.4519 & 0.5305 & 0.5243\\
Llama-3  &8,030&
0.4904 & 0.6412 & 0.5577 & 0.5191 & 0.5267 & 0.7690 & 0.5454 & 0.5342 & 0.5729\\
Mistral  &7,248& 
0.4833& \textbf{0.6933}& 0.559& 0.5321& 0.5313& 0.8102& 0.5762& 0.5677
& 0.5941\\
 Deepseek-Qwen2& 7,615& 0.5117& 0.5407
& 0.563& 0.5165& 0.5303& 0.5905& 0.5994& \textbf{0.5648}&0.5520\\
GPT-3.5  &-& 
0.5057 & 0.5170 & 0.5110 & 0.5122 & 0.5039 & 0.6184 & 0.5801 & 0.5076 & 0.5319\\
\midrule
RecBase$_\text{base}$ &313& 
\textbf{0.5508}& 0.5352& 0.5401& 0.5029& 0.5320& 0.7450& 0.5870& 0.4874& 0.5601\\
RecBase$_\text{large}$ &1,318& 0.5442& 0.6474& \textbf{0.5712}& \textbf{0.5329}& 0.5326& \textbf{0.8343}& \textbf{0.6761}& 0.5124&\textbf{0.6063}\\
\bottomrule

\end{tabular}
}
\end{table*}

\section{Experiments}

\begin{figure}[h]
    \centering
    \includegraphics[width=\columnwidth]{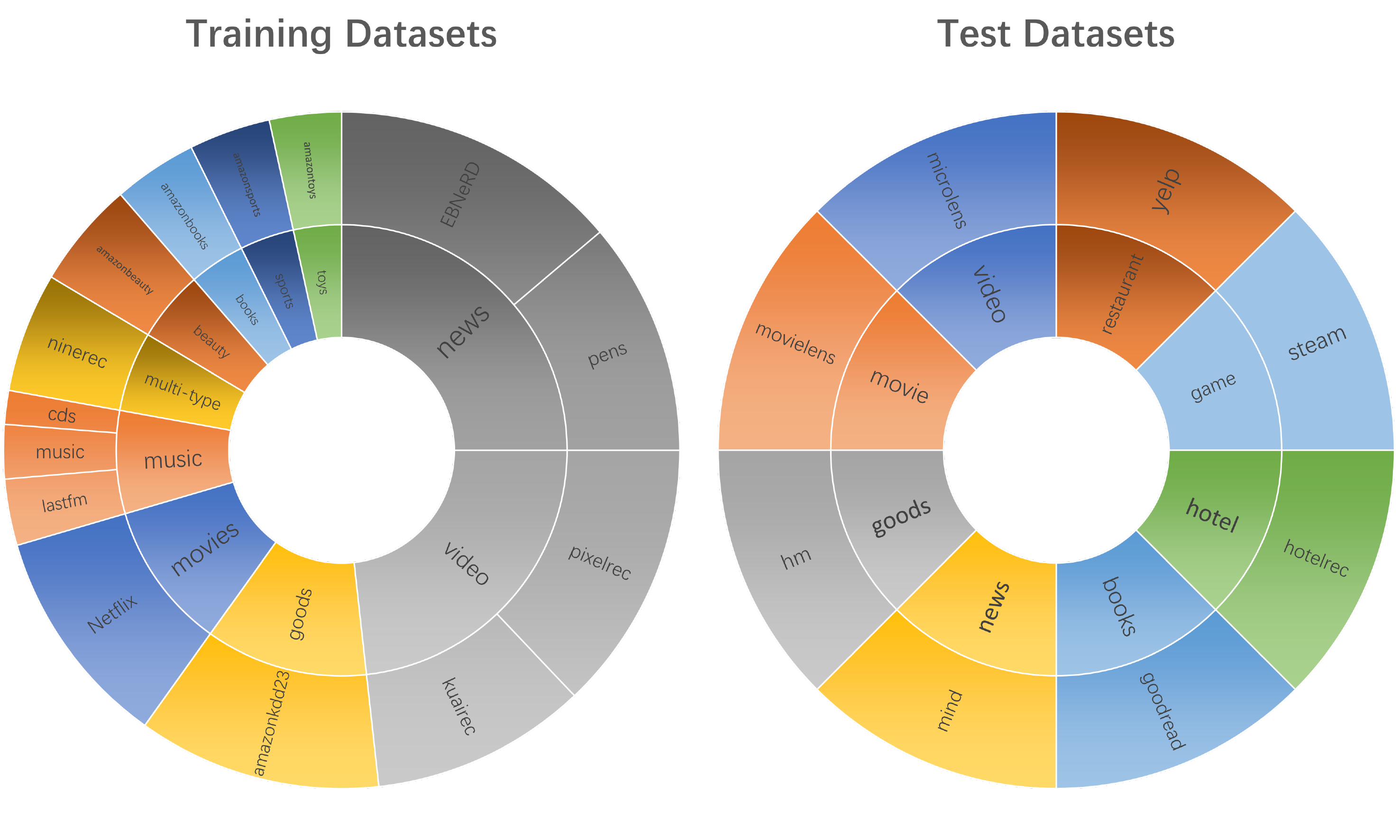}  
    \caption{Training and test datasets distribution}  
    \label{fig:dataset}  
\end{figure}

We evaluate our method through zero-shot recommendation experiments, pre-training our model on 15 diverse datasets and testing it on 8 unseen datasets to assess its generalization ability. The dataset statistics are shown in Figure~\ref{fig:dataset} and Table~\ref{statistic}.

\subsection{Evaluation Setup}

\textit{RecBench}~\cite{recbench,recbench-md} is a recently introduced benchmark (comprising 15 datasets across 10 domains) designed to evaluate LLM-based recommenders. In this work, we adopt RecBench for zero-shot recommendation evaluation in the majority of our experiments. More specifically, we use item textual information to represent each item, including that in the user browsing history. Given a user--item pair, we will use natural language to concatenate the user and item feature:
``The user has browsed the following items: ... , will this user be interested in the item: ...? Answer (Yes/No): ''. The prediction of the next token will be regarded as the user interest score or click probability). For closed-source models, we map the LLM’s textual responses (\text{YES} or \text{NO}) directly to interest scores of 1.0 and 0.0, respectively. For open-source models, we obtain the logits of the \text{YES} and \text{NO} tokens from the classifier, denoted as $l_{\text{yes}}$ and $l_{\text{no}}$. After applying softmax normalization \textbf{over these two tokens}, we take the score corresponding to the \texttt{YES} token as the click probability, formulated as:
\begin{equation}
\text{Click Probability} = p_{\text{yes}} = \frac{e^{l_{\text{yes}}}}{e^{l_{\text{yes}}} + e^{l_{\text{no}}}}.
\end{equation}

For inference, RecBase takes the user's historical interaction sequence as input and outputs the logits for the predicted item concept IDs. These logits represent the joint probability distribution over the \( m \) possible concept IDs. By comparing these probabilities, we identify the item that the user is most likely to engage with, based on their previous interactions. This process allows RecBase to generate recommendations by ranking items according to their predicted interest scores.

The task is framed as predicting user interest in unseen items based on historical interactions, approached as a ranking problem. To evaluate the model’s performance, we use the \textbf{Area Under Curve (AUC)} metric, which measures the model’s ability to rank items in order of relevance.

\begin{table}[ht]
\centering
\renewcommand{\arraystretch}{1.2} 
\caption{Statistics of training, finetune, and test datasets.}

\resizebox{0.48\textwidth}{!}{
\begin{tabular}{clccccccccc}
\toprule
    & Item size & User size & History Avg. length\\
\midrule
Training datasets  & 4,595,003 & 35,047,682 & 20.37 \\
Finetune datasets  & 1,005,745 & 5,098,084 & 17.83 \\
Test datasets  & 623,615 & 145,975 & 15.01 \\
\bottomrule
\end{tabular}
}
\label{statistic}
\end{table}

\subsection{Implementation Details}

In our approach, we employ a hierarchical feature extraction technique using the CL-VAE model with a 4-level codebook, each level sized at 2048. This setup enables the model to progressively extract more structured features from raw data, as validated by ablation studies. For textual item descriptions, we use the NV-Embed-v2 \cite{NV-emb} model to convert unstructured text into dense, semantically rich embeddings, making the data suitable for downstream recommendation tasks.

For the RecBase pre-training, we train two versions: Base and Large. The Base version is configured with a hidden size of 1024, an intermediate size of 2816, and 16 attention heads across 24 layers. It supports a maximum position embedding length of 32,768. The Large version increases the hidden size to 1536, intermediate size to 8960, and adjusts the number of attention heads to 12 with 28 layers. The position embedding length and sliding window are extended to 131,072, enhancing the model's capacity to process longer sequences. Both models use a vocabulary size of 20,000 and share other key settings derived from the Qwen2 \cite{Qwen2-tech} architecture.

\subsection{Zero-Shot Recommendation Performance}
In this experiment, we evaluate our RecBase model against several state-of-the-art approaches, including zero-shot recommendation methods based on large language models (LLMs) and fine-tuned recommendation models. Specifically, we compare our model with LLM-based zero-shot methods such as BERT\textsubscript{base} \cite{bert}, OPT \cite{opt}, Qwen-2 \cite{Qwen2-tech}, Phi-2, Llama-2 \cite{llama2}, Llama-3 \cite{llama3}, Mistral \cite{mistral}, and GPT-3.5 \cite{GPT3}, as well as fine-tuned LLM-based models like RecGPT \cite{recgpt} and P5 \cite{P5}. This comparison is conducted across multiple datasets to evaluate the generalization ability of each method.

The experimental results demonstrate the superiority of RecBase over traditional LLM-based methods in both generalization and efficiency. Specifically, RecBase$\text{large}$ achieves an overall score of 0.6063, surpassing strong baselines such as RecGPT, P5, Mistral, and GPT-3.5. These findings highlight the effectiveness of our domain-specific pretraining strategy, which enables the model to capture fine-grained semantic nuances of recommended items and achieve robust zero-shot performance. Moreover, RecBase consistently outperforms advanced LLM-based models such as Llama-3 and Qwen-2 across all datasets, with particularly notable improvements on H\&M (0.6761 vs. 0.6287) and Steam (0.8343 vs. 0.8102), underscoring its strong generalization capability. In addition, RecBase$\text{base}$, with only 313M parameters, delivers competitive results while outperforming models like BERT$\text{base}$ and OPT$\text{base}$, all at a substantially lower computational cost. Collectively, these results position RecBase as both a highly efficient and powerful solution for zero-shot recommendation ranking.

\begin{table}[t]
\centering
\renewcommand{\arraystretch}{1.4}

\caption{Performance of our model under zero-shot and fine-tuning settings on various datasets.}

\resizebox{\columnwidth}{!}{
\begin{tabular}{cccccccccc}
\toprule
    & Microlens & Steam & MovieLens & H\&M & Yelp \\
\midrule
Zero-shot & 0.5401 & 0.7450 & 0.5352 & 0.5870 & 0.5320  \\
Fine-tuned   & 0.5602 & 0.9173 & 0.6216 & 0.6261 & 0.6125  \\
\midrule
Improve. (\%)  & \textbf{3.70\%} & \textbf{23.12\%} & \textbf{16.14\%} & \textbf{6.66\%} & \textbf{15.13\%}  \\
\bottomrule
\end{tabular}
}
\label{tab:zero-fine}
\end{table}

\subsection{Unified Representation Performance}
To evaluate the effectiveness of the unified concept space generated by the CL-VAE method, we conducted an in-depth analysis. As previously noted, an optimal code discretization approach should aim to map the input data distribution as uniformly as possible into the latent space, maximizing the utilization of each token. This ensures that the autoregressive model can effectively leverage every token value for encoding. As illustrated in Figure~\ref{fig:clustering}, traditional methods such as RQ-VAE result in features from different domains being independently distributed in the latent space, with minimal interaction between them. In contrast, the CL-VAE method significantly improves upon this, as evidenced by the increased overlap and interaction between features from diverse datasets within the unified concept space.

\begin{figure}[htbp]
    \centering
    \begin{subfigure}{0.49\columnwidth} 
        \centering
        \includegraphics[width=\columnwidth]{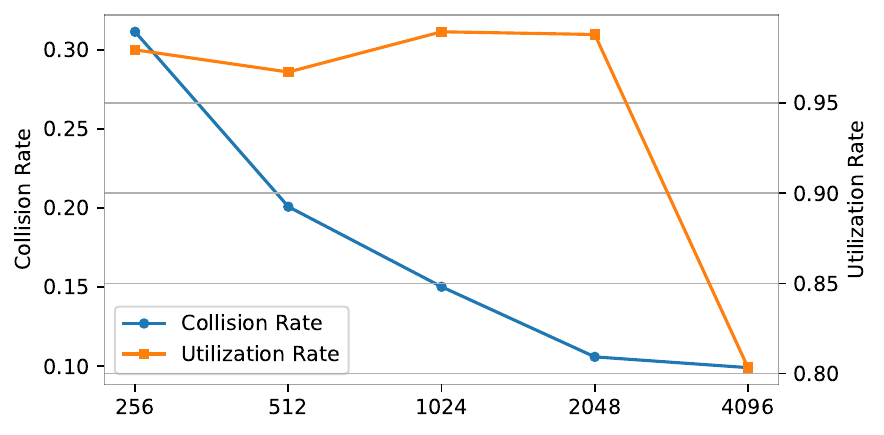}
        \caption{codebook size}
        \label{fig:cluster1}
    \end{subfigure}
    \hfill
    \begin{subfigure}{0.49\columnwidth} 
        \centering
        \includegraphics[width=\columnwidth]{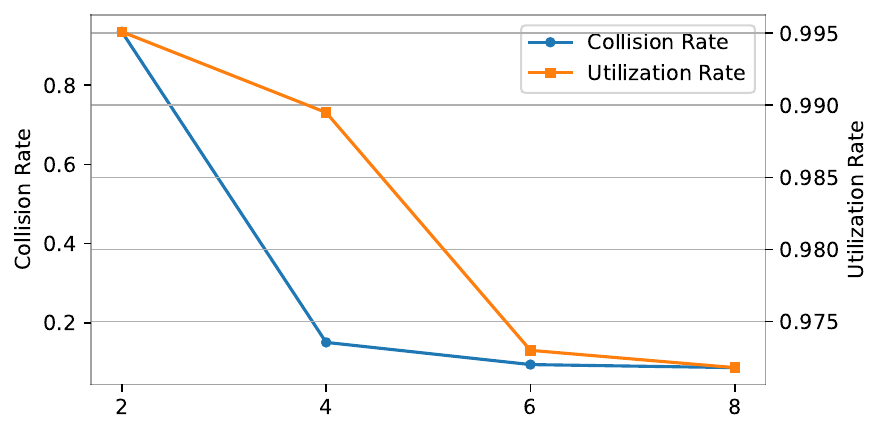}
        \caption{codebook level}
        \label{fig:cluster2}
    \end{subfigure}
    \vfill 

    \begin{subfigure}{0.95\columnwidth} 
        \centering
        \includegraphics[width=\columnwidth]{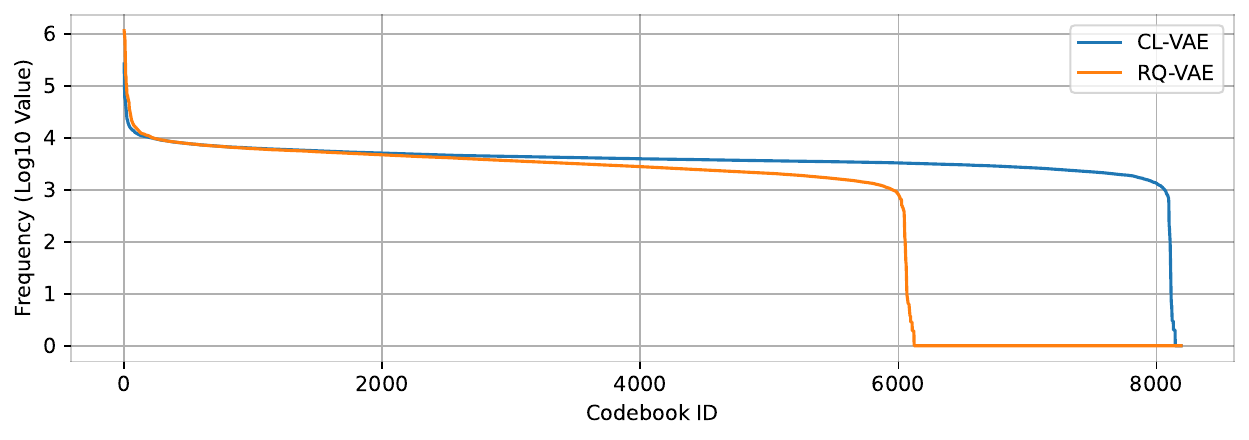}
        \caption{Codebook usage frequency in CL-VAE and RQ-VAE. }
        \label{fig:codeusage}
    \end{subfigure}

    \caption{Codebook structure vs. Collision rate and Utilization rate. (a) demonstrates the impact of codebook size on the collision rate and utilization rate. (b) reflects the influence of codebook level on the aforementioned metrics.}
    
    \label{fig:codebook}
\end{figure}

The transformation in the frequency distribution of code usage across different levels, as depicted in Figure~\ref{fig:codeusage}, further supports the validity of this mapping. The CL-VAE method not only achieves a more balanced distribution of IDs but also ensures the effective exploitation of the hierarchical structure of the codebook. This hierarchical approach enables the model to capture both fine-grained and coarse-grained features, thereby enhancing its generalization capability across various recommendation scenarios. By establishing a unified and well-distributed concept space, CL-VAE facilitates the autoregressive model in making more efficient and accurate predictions, ultimately improving the overall performance of the recommendation system.

\section{Analysis}
In this section, we present a series of ablation studies to evaluate the contribution of individual components to the model’s overall performance and generalization ability.

\subsection{Ablation Analysis on Key Components}
The ablation study, as shown in Table~\ref{ablation}, highlights the importance of each component within CL-VAE. The baseline model, RecBase$_\text{base}$, achieves the best performance across all datasets, demonstrating the effectiveness of the complete method. When the formatted text description is removed, there is a noticeable drop in performance, indicating that structured text representations play a vital role in enhancing the model's ability to capture relevant features. Similarly, the removal of the reinitialization step results in a significant decline in performance, suggesting that the initialization mechanism is crucial for stabilizing learning and improving convergence. Additionally, excluding the curriculum learning module leads to further performance degradation, particularly in more complex recommendation scenarios, which underscores the value of progressively training the model on increasingly difficult examples. 

\begin{table}[t]
\centering
\renewcommand{\arraystretch}{1.2} 
\caption{Modular ablation study. format., init. and cur. represent formatted text description, reinitialization and curriculum learning in CL-VAE respectively.}

\resizebox{0.48\textwidth}{!}{
\begin{tabular}{ccccccccccc}
\toprule
    & Yelp & Steam & H\&M & HotelRec & Overall \\
\midrule
RecBase$_\text{base}$  & 0.5320 & 0.7450 & 0.5870 & 0.4874 &0.5879 \\

\midrule

w/o format.  & 0.5204 & 0.7187 & 0.5668 & 0.4966 &0.5756 \\
w/o init.  & 0.4912 & 0.5924 & 0.5319 & 0.4909 &0.5266\\
w/o cur.  & 0.5073 & 0.6815 & 0.5412 & 0. 4815 &0.5529 \\
\bottomrule
\end{tabular}
}
\label{ablation}
\end{table}

\subsection{ Ablation Study on the Codebook }
We perform an ablation analysis on the size and number of levels in the multi-level codebook of the CL-VAE module. As shown in Figure 2, with an increase in size, the conflict rate between the concept IDs obtained from the codebook continuously rises, indicating that enlarging the size benefits the optimization of representations. However, when the size reaches a certain scale, such as 4096 in the figure, a decrease in utilization occurs, leading to redundant wastage of the vocabulary space. Therefore, our model selects a size of 2048 for each layer. Regarding the analysis of levels, as the number of levels increases, the number of products that can be represented by the concept IDs grows exponentially. For spaces beyond four levels, the utilization of IDs becomes very low, and the gains from increasing the number of levels begin to plateau, further incurring additional inference costs during decoding. Consequently, our model adopts a strategy of four levels to optimize the balance between performance and efficiency.

\subsection{In-Domain Adaptation via Fine-Tuning}
Table \ref{tab:zero-fine} shows the performance of our model under zero-shot and fine-tuning settings across various datasets. Fine-tuning consistently improves performance over the zero-shot setting, with significant gains on the Steam (+0.1723) and MovieLens (+0.0864) datasets, indicating the effectiveness of domain-specific adaptation. Even datasets with smaller improvements, such as Microlens (+0.0201) and Yelp (+0.0805), benefit from fine-tuning. Figure \ref{fig:pipe} illustrates the fine-tuning process, emphasizing how in-domain adaptation refines the model’s representations. These results demonstrate the adaptability and potential of our model for further performance enhancement with in-domain supervision.

\subsection{Analysis of Inference Efficiency}
We conducted an inference efficiency comparison between our model and several other state-of-the-art models, including Qwen2, phi, GptRec, Mistral, and RecBase, on the same experimental dataset. Our model consistently demonstrated superior inference efficiency, outperforming the others by a significant margin. This improvement can be attributed to the specialized ID vocabulary space we designed specifically for recommendation tasks. Unlike general-purpose large language models that rely on vast vocabulary spaces based on natural language representations, our model utilizes a much smaller, more efficient vocabulary tailored to the needs of recommendation systems. This design choice not only enhances the model's efficiency but also positions it as a more suitable base model for recommendation-related tasks.

\begin{figure}[h]
    \centering
    \includegraphics[width=\columnwidth]{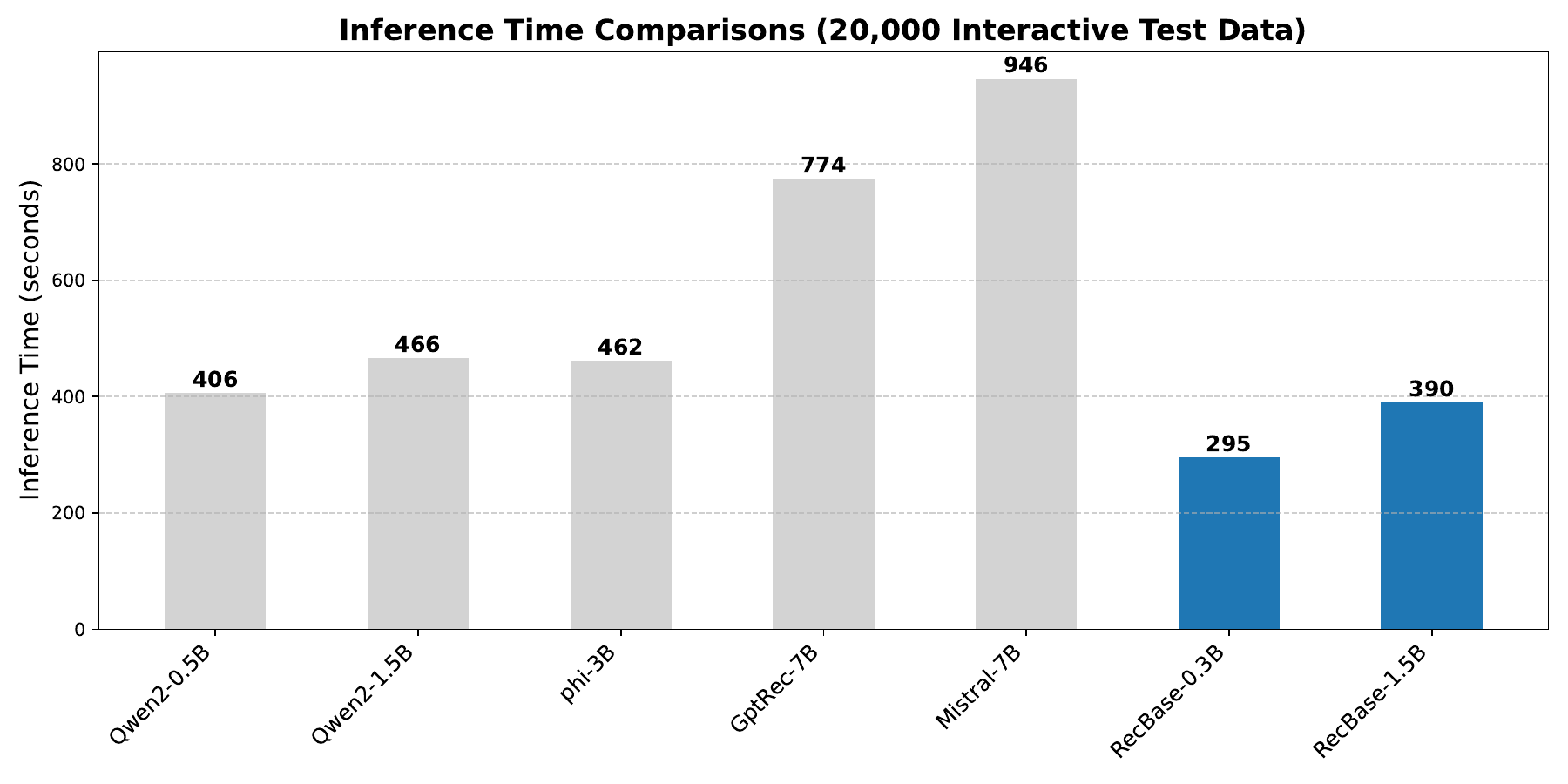}  
    \caption{Comparison of Inference Latency.}  
    \label{fig:line}  
\end{figure}

\section{Conclusion}

In this paper, we introduce RecBase, a foundation model tailored to the challenges of zero-shot and multi-domain recommendation. By pretraining on a large-scale cross-domain corpus with structured text representations and unified feature mappings, RecBase demonstrates strong generalization across heterogeneous recommendation tasks. The incorporation of curriculum learning and discrete representations facilitates the construction of a unified concept ID space, thereby mitigating semantic discrepancies between domains. Furthermore, the autoregressive training paradigm enables the model to effectively capture inter-item dependencies, yielding superior zero-shot and cross-domain performance compared to conventional large language models. Comprehensive evaluation on eight real-world datasets confirms the effectiveness of our approach, particularly in cold-start scenarios, highlighting the potential of recommendation-oriented pretraining as a promising direction for building robust and adaptable recommender systems.

\section{Limitation}

Despite its promising performance in zero-shot and multi-domain scenarios, our model exhibits several limitations inherent to recommendation data. Data sparsity and distribution imbalance can impair generalization, particularly for cold-start users and long-tail items. While cross-domain pretraining partially alleviates this, the model may still underrepresent certain domains or items with sparse interactions. Additionally, biases in the training data can limit generalization to new domains or diverse user populations. Future work should explore data augmentation, active learning, and bias mitigation strategies, and evaluate the model on larger, more heterogeneous benchmarks to enhance scalability and real-world robustness.

\section*{Acknowledgement}
This work was supported by the National Key R\&D Program of China (2022ZD0162000) and the National Natural Science Foundation of China under Grants No.62222211 and No.U24A20326 . We also acknowledge partial support from MindSpore (\url{https://www.mindspore.cn}), a new deep learning computing framework.

\bibliography{custom}

\section*{Supplementary Material}

\appendix

\section{Dateset Details.}
\label{sec:appendix}

The rec-base model are pretrained on 15 diverse training datasets across various domains and evaluated on 8 additional cross-domain datasets. Here are detail descriptions about these datasets.

\paragraph{Training datasets} EBNeRD \cite{EBNeRD} and PENS \cite{PENS} are news-related datasets. EBNeRD is used to optimize news recommendation systems, while PENS is a dataset for personalized news headline generation. 
The PixelRec \cite{pixelrec} and KuaiRec \cite{kuairec} datasets are both related to short video recommendations, containing a large number of user interactions with short videos and the corresponding video thumbnails. The Amazon Reviews 2023 \cite{Amazonreview} dataset is a large-scale e-commerce review dataset that covers user feedback information such as ratings, review texts, and helpful votes, as well as product metadata. The Amazon Review Dataset records user evaluations of products on the Amazon website and is a classic dataset for recommendation systems. Sub-datasets such as Amazonbeauty, Amazonbooks, Amazonsports, and Amazontoys focus on the fields of beauty products, books, sports products, and toy products, respectively. Netflix dataset is a classic movie-related recommendation system dataset, containing over 100 million user ratings for movies. 

\paragraph{Evaluation datasets} MIND \cite{mind} dataset is a large-scale news recommendation dataset constructed from Microsoft News user click logs, featuring rich information for each news article, including title, abstract, body, and category labels. It records users' news click history and impression logs. The dataset is well-suited for studying challenges in news recommendation such as cold start problems, and user interest modeling. MovieLens \cite{movielens} dataset is a classic recommendation system dataset created by the GroupLens research team at the University of Minnesota, containing a large number of movie ratings by users. MicroLens \cite{microlens} is a large-scale, content-driven short video recommendation dataset that contains 1 billion user interactions with short videos and provides rich modality information for these videos. The Goodreads \cite{goodreads} dataset was collected from the Goodreads website by UC San Diego, containing book metadata, user-book interactions, and detailed user reviews, with approximately 228 million user-book interaction records. The Yelp Open Dataset is a subset of business, review, and user data provided by the well-known American merchant review website Yelp. It includes approximately 160,000 businesses, 8.63 million reviews, and 200,000 images from eight major metropolitan areas. The Steam Dataset is a multi-dimensional dataset built based on Steam, the world's largest digital game distribution platform. It covers detailed information such as game purchase records and playtime for millions of users and is widely used for research in user behavior analysis, market trend prediction, and game recommendation systems. The H\&M dataset is a dataset provided by H\&M, containing product information, customer information, and transaction records. It is widely used for research in recommendation systems.  HotelRec \cite{hotelrec} is a large-scale hotel recommendation dataset created by the Artificial Intelligence Laboratory at École Polytechnique Fédérale de Lausanne (EPFL), collected from the TripAdvisor platform. It contains approximately 50 million hotel reviews and is the largest recommendation dataset in a single domain with text reviews.

\section{ Data Processing Details}

\subsection{ Biases in Data Distribution and Modality Handling}

The RecBase pretraining corpus is constructed from 15 cross-domain recommendation datasets. We acknowledge certain biases in the overall distribution, particularly the overrepresentation of news and audio-visual content. To mitigate this, we carefully curated the dataset selection to achieve a more balanced category distribution, ensuring that no single domain dominates the training signal.

To ensure a controlled and fair evaluation, we standardized all inputs to a text-only format. For multimodal datasets such as PixelRec, Clothing, and NineRec, we therefore isolated and used only their textual descriptions (e.g., product metadata and reviews), while discarding other modalities. This design choice mitigates confounding effects from heterogeneous data types and establishes a level playing field for comparison with our text-centric baselines. While this precludes the use of multimodal signals for now, we argue the rich textual information present in these datasets provides a robust basis for assessing model generalization. The integration of discretized multimodal features into our unified framework is a promising direction for future research.

\subsection{Data Preprocessing and Noise Filtering}

To maximize exposure to real-world user behavior patterns, RecBase is pretrained on raw user interaction data from each dataset. The preprocessing pipeline includes the following steps:

\paragraph{Text Standardization.} We structured item-related content into a unified format, combining titles, attributes, and reviews into a clean textual description. Non-informative or off-topic reviews were filtered out to retain only content relevant to the core product or item, as illustrated in Figure~1(b) and detailed in Appendix~B.

\paragraph{User History Filtering.} We applied length-based filtering to manage the variance in user history lengths. Users with fewer than 15 interactions (e.g., in \textit{AmazonToys}) were removed due to insufficient sequence signal, while extremely long histories (e.g., exceeding 2500 interactions in \textit{LastFM} or \textit{PixelRec}) were truncated to avoid overfitting and memory inefficiency.

\subsection{ Negative Sampling}

Unlike traditional recommendation models that require explicit negative sampling, RecBase is trained on real user interaction sequences using an autoregressive objective. Therefore, no artificial negative sampling is involved during training. The model learns to predict the next likely item based on historical sequences, leveraging only positive (i.e., observed) user feedback. This setup reflects the natural sequential structure of user behavior and aligns with the pretraining objective.

\section{Examples of Items.}

In what follows, we show some examples of formatted text descriptions of items for extracting embeddings by using the NV-Embedding model~\cite{NV-emb}:
\begin{tcolorbox}[colback=gray!10,
                  colframe=black,
                  width=0.48\textwidth,
                  arc=1mm, auto outer arc,
                  boxrule=0.5pt,
                  box align=center,
                 ]
'Describe a movie:\textbackslash n\{\textbackslash n"title": "Toy Story (1995)",\textbackslash n"genres": "Adventure | Animation | Children | Comedy | Fantasy"\textbackslash n\}'
\\
\\
'Describe a movie:\textbackslash n\{\textbackslash n"title": "Jumanji (1995)",\textbackslash n"genres": "Adventure | Children | Fantasy"\textbackslash n\}'
\\
\\
'Describe a movie:\textbackslash n\{\textbackslash n"title": "Grumpier Old Men (1995)",\textbackslash n"genres": "Comedy | Romance"\textbackslash n\}'
                 
\end{tcolorbox}

Test cases for large language model benchmarks:
\begin{tcolorbox}[colback=gray!10,
                  colframe=black,
                  width=0.48\textwidth,
                  arc=1mm, auto outer arc,
                  boxrule=0.5pt,
                  box align=center,
                 ]
'User behavior sequence: \textbackslash n(1) This Ford GT40 Movie Rig From "Ford V Ferrari" Looks Absurd\textbackslash n(2) Kendall Jenner Wore the Tiniest Dress to Go Jewelry Shopping\textbackslash nCandidate item: 9 fashion trends inspired by the 2000s that are coming back in style'
\\
\\
'User behavior sequence: \textbackslash n(1) This Ford GT40 Movie Rig From "Ford V Ferrari" Looks Absurd\textbackslash n(2) Kendall Jenner Wore the Tiniest Dress to Go Jewelry Shopping\textbackslash nCandidate item: Here Are the Biggest Deals We\'re Anticipating for Black Friday'
\\
\\
'User behavior sequence: \textbackslash n(1) This Ford GT40 Movie Rig From "Ford V Ferrari" Looks Absurd\textbackslash n(2) Kendall Jenner Wore the Tiniest Dress to Go Jewelry Shopping\textbackslash nCandidate item: Man cuffed for eating sandwich on train platform gets an apology'
\end{tcolorbox}

\end{document}